\documentclass[a4paper,11pt]{article}

\usepackage[latin1]{inputenc}
\usepackage{graphicx}

\def\vareps{  {\varepsilon} }
\def\interi{\mathinner{\bf Z}}
\def\reali{\mathinner{\bf R}}
\def\fraz#1#2{ {#1 \over #2} }
\def\vett#1{ {\mathbf{ #1}} }
\def\tdot#1{\hskip2pt\ddot{\null}\hskip2.5pt \dot{\null}\kern -5pt {#1}}

\title{Nonradiating normal modes in a classical many--body  model of
matter--radiation interaction}
\author{A. Carati\footnote{ Universit\`a di Milano, Dipartimento di Matematica
Via Saldini 50,  20133 Milano (Italy)  E-mail: {\tt
carati@mat.unimi.it} }\,   and L. Galgani\footnote{ Universit\`a di Milano,
Dipartimento di Matematica, Via Saldini 50,  20133 Milano (Italy)  
E-mail: {\tt galgani@mat.unimi.it} } }

\begin{document}

\maketitle

%\vskip 1 truemm
%\centerline{ Universit\`a di Milano, Dipartimento di Matematica }
%\vskip 1 truemm
%\centerline{Via Saldini 50,  20133 Milano (Italy)}
%\vskip 1 truemm
%\centerline{ E-mail: {\tt carati@mat.unimi.it} }
%\vskip 1 truecm

\centerline {ABSTRACT} 
\noindent  
\vskip 1truecm
We consider a classical model of matter--radiation  interaction,
in which the matter is represented by a
system of infinitely many dipoles on a one--dimensional
lattice, and the system is dealt with in the so--called dipole (i.e. linearized)
approximation.   We prove that there exist normal--mode solutions of the
complete system, so that in particular the dipoles, though performing
accelerated motions,  do not radiate energy away. This comes about 
in virtue of an exact compensation which we prove to occur, for each
dipole, between the  ``radiation reaction force'' and a part of the
retarded forces due to all the other dipoles. This fact corresponds to
a certain identity which we name after Oseen, since it occurs that 
this  researcher  did actually propose it, already in the year 1916.
We finally make a connection with a paper of Wheeler and
Feynman on the foundations of electrodynamics. It turns out indeed that the
Oseen identity, which we prove here in a particular model,  is in fact a weak
form of a general identity that such authors were assuming as an independent
postulate.
\vfill  %%% DA METTERER IN FONDO 
%\noindent PACS numbers: 03.50.D, 03.65.B,  02.30.M 
%\par

%%% DA CANCELLARE
\noindent
Running title: normal modes in matter-radiation interaction  
\vskip 3.truecm
\noindent
\eject %%%% Da cancellare se non si vuole l'abstract in una pagina

\section{Introduction} 
In the framework of classical electrodynamics it is often given for
granted that a charged particle, when accelerated under the influence
of some external force, gives off radiation and thus loses energy.
Actually, if a single particle is considered, this is only partially
true. Indeed, if the contribution of the radiative interaction to the
equation of motion is taken into account in the standard way through
the familiar ``radiation reaction force''  proportional to the time
derivative of the acceleration (or through its relativitic extension
proposed by Dirac \cite{dirac}), then a harmonic oscillator is easily proven to
steadily lose energy and fall onto the center of attraction, while in
the case of a Coulomb attraction the particle is found to start losing
energy but finally to escape to infinity \cite{eliezer}. 
Anyway, in such two cases
involving a single charged particle it has been proven that
oscillatory motions do not exist.

In the present paper we investigate the existence of oscillatory
motions in the case of $N>1$ charged particles.  In fact, we even consider
a system of infinitely many particles, for an extremely simplified
model particularly suited for an analytical investigation. This is a
system of equal particles (which we also call ``dipoles'' or `` linear
resonators''), that are all constrained to move on a line and that, in
the absence of any electrodynamical interaction, would perform linear 
oscillations each
about a proper site, the sites constituting a periodic lattice on the
given line.  The electrodynamic interaction is introduced in the
standard way which was fixed by Dirac for a general system of particles and field,
taking however  the so--called ``dipole
approximation'' (i.e. linearizing the equations of motion with respect
to the displacements of the resonators, and to their time derivatives).
 In such a way, by the standard procedure of eliminating the field in
the coupled equations for matter and field,  one
is reduced to a system of equations which are just the mechanical
Newton equations for each dipole, in  which the effect of the
electromagnetic interaction appears through two contributions,
namely: the familiar  ``radiation reaction force'' on each dipole (in
the dipole approximation, which is here equivalent to the
nonrelativistic approximation), and a  mutual retarded force
between each pair of dipoles.  An  interaction with an external free field
could also have been considered. But we drop it, because it mainly plays a
role in connection with dispersion theory, with which we are not  directly
concerned here. The equations of motion of the model
are  written down below (see
(\ref{modellovecchio})).  In fact, apart from the special choice of
the disposition of the equilibrium positions of the resonators, the
present model is nothing but the standard one that is usually
employed for a microscopic molecular foundation of optics.
Actually, this is only partially true, because the necessity of
introducing the radiation reaction force in such models has been from
time to time put in doubt. Here, not only do  we  introduce
such a  radiation reaction term  into the model, relying on the authority
of Dirac, but also claim that its role is clarified by
the main result of the present paper.

Indeed we prove the rather surprising  result that there exist solutions of the
complete system of equations, in which the retarded electromagnetic forces
produced on a given dipole  by the fields ``created''  by    all the other
ones add up in such a way as to  exactly compensate the reaction
radiation force acting on it, so that one remains with no radiation at
all. The relevant point is that such nonradiating solutions, which
have the form of normal
modes for the complete system, could not exist if the radiation
reaction term had not been included in the model.  
Presumably,  the present result  might prove useful in
establishing the existence of collective normal--mode  motions also  in a quantum
mechanical version of the present model of matter--radiation
interaction.

The way we conceived that nonradiating normal--mode solutions should
exist for models of matter--radiation interaction is the following
one. We were interested in the models of matter--radiation interaction
that Planck had actually been considering as mimicking a black
body. We thus found out that his models, although in principle
involving $N$ resonators, were actually dealing with a single
resonator acted upon by an external field: indeed Planck explicitly
made the assumption that the $N$ resonators act `` independently of
one another''. At first sight, such an assumption might appear just an
innocuous one, perhaps consituting an excessive simplification of the
physical problem, but for the rest acceptable. Things changed however
when we suddenly realized \cite{guerra} that such a simplified model is actually
inconsistent. Indeed, if the resonators are supposed to act
incoherently, then, at a given point, the far fields radiated by each
of them are easily seen to add up to give a divergent contribution (or
a contribution proportional to the volume in the case of finite $N$),
so that one meets here with a paradox analogous to that of Olbers.
Thus it seemed to us that the inconsistency could be removed only if
the system of resonators coupled by radiation  were considered in its
globality, and special solutions were  looked for, which present a
coherent character, i.e. are in the form of normal modes. Indeed, with
suitable coherent motions one might obtain that the multipoles of
higher and higher orders are made to vanish as $N$ is increased, so
that the total  emitted field  would vanish in the limit $N\to\infty$
(see \cite{landau}).  These two arguments agree in indicating that a large number
of particles should be considered if normal modes have to be looked for. The
choice of taking an infinite number of them was just made in order to
obtain  a simplification in the mathematical treatment of the problem,
as occurs in so called ``thermodynamic limit'' of statistical
mechanics. This is the way we happened to find out the existence of
normal modes. The core of the result is the identity mentioned above,
which gives an exact compensation (or cancellation) of the reaction
radiation term of each dipole, by a part of the sum of  the retarded
fields produced by all the other dipoles.

Having found such a result, we started a bibliographical research with
the aim of learning whether something analogous was already
known. Through the book of Born and Wolf \cite{bornwolf} we went back to the
celebrated papers of Ewald \cite{ewald} and Oseen \cite{oseen1}.  
The best source of
information  proved  however   to be the long review written by
Jaff\`e \cite{jaffe} for 
the Handbuch der Experimentalphysik. So we learned first of all
that the kind of model considered by us (dipoles with mutual retarded
electromagnetic interactions) was completely standard (apart from
discussions about the feasibility of the reaction radiation term) in
the many studies on the microscopic molecular foundation of optics,
that were filling up a consistent part of the issues of Annalen der
Physik in the years 1915--1925.  It should be mentioned however that
essentially all such papers (with the exception of that of Ewald,
which on the other hand apparently might present serious drawbacks) actually had a
somehow mixed character.  Indeed, in estimating the total force acting
on a single resonator due to all the other ones, use was made of an
approximation in which the relevant sums were replaced by appropriate
integrals, somehow in the spirit of continuum mechanics. Thus the
problem could not be dealt with in a mathematically clearcut way, as
occurs with  a theorem for a system of differential equations.  From the review
of Jaff\`e we also learned that our main identity producing the
mentioned cancellation had already been introduced by Oseen
\cite{oseen2} in a
paper subsequent to the one mentioned by Born and Wolf. The proof was
given however in the spirit quoted above of continuum mechanics, so
that in particular it was not clear 
whether the identity applies only to crystals or also to gases, or
even whether it holds at all.  Furthermore it occured that Oseen was
making use of his result in a critique \cite{oseen3} to a previous theory of
Planck on the dispersion of light.  A long debate followed, in which the two
questions, truth of the identity and soundness  of Planck's dispersion
theory, were accomunated. The final conclusion, explicitly stated in
the review of Jaff\`e (see page 266), 
was that the theory of Oseen was actually wrong
(\emph{irrig}).  As a consequence, it occurred that the identity
itself was apparently discarded by the scientific community, and
eventually forgotten.  Because of all these considerations concerning
the identity in question which we like to call the Oseen identity
(i.e. lack of an explicit proof in a concrete model of differential
equations, and its having been discarded by the scientific community),
we came to the conclusion that our proof, produced as a theorem for a
concrete model, might be worth of publication.

A further point of possible interest of our result is its strong  relation
with the paper of Wheeler and Feynman \cite{wf} on the foundations of
electrodynamics. Indeed it turns out that the Oseen identity, which is
here proven in our model, is nothing but a weak form of an identity
that  plays a central role in the paper of Wheeler and Feynman,
and is by them assumed as an independent postulate.

In Section 2 the model is defined. In Section 3 the Oseen identity and
the existence of normal modes are proven, while in Section 4 
the form of the dispersion relation is discussed. In
Section 5 the Oseen identity is expressed in terms of the forces
acting on the dipoles, while  the connection with the paper of Wheeler
and Feynman is  discussed in Section 6. Some conclusive comments follow in
Section 7.

\section{The model} We consider an infinite number
of charged particles of equal mass $m$ and charge $e$ constrained on a
line, and denote by
 $x_j$, with $j\in\interi$, their cartesian coordinates. We assume there exists
an equilibrium configuration with positions  $x_j=r_j$, where
$r_j=ja$ and  $a$ is a positive parameter (the lattice step), 
corresponding to  a balance of
the mutual Coulomb forces and of other possible  mechanical forces.
In the absence of further electrodynamical interactions, we assume
that each particle performs a linear oscillation
with the same  charactereristic  angular frequency  $\omega_0$
about its equilibrium position $r_j$. The interaction with the
electromagnetic field  is
taken into account in the dipole
approximations as described below, to the effect that the final
mathematical model has as unknowns only the displacements
$q_j=x_j-r_j$ of the particles from their equilibrium positions, and
that such displacements satisfy the infinite system of delayed differential
equations (for $j\in\interi$)
\begin{equation}\label{modellovecchio}
m (\ddot q_j + \omega_0^2 q_j -\vareps \tdot q_j ) = 2e^2  \sum_{k\ne
        j} \bigg[ \fraz {q_k(t-r_{jk}/c)}{r_{jk}^3} + \fraz 1c \fraz {\dot
        q_k(t-r_{jk}/c)}{r_{jk}^2}\bigg] \ ,
\end{equation}
where the sum is extended over $k\in\interi$, $k\ne j$. Here 
 $r_{jk}=|r_j-r_k|$ is  the distance between particles $k$ and
$j$, evaluated at their equilibrium positions,  $c$ is the speed of
light, and the familiar parameter
$$
\vareps= \fraz 23 \fraz {e^2}{mc^3}
$$
has been introduced.   As mentioned in
the Introduction, these equations, apart from the special choice of
the equilibrium positions of the particles, are the ones that were 
commonly used for a
molecular foundation of optics in the years 1915--1925. Actually, this
is completely true only for what concerns the right hand sides of the
equations, because a general agreement had not yet been reached
concerning the feasibility of the radiation reaction term
$-m\varepsilon \tdot q_j$ appearing at the l.h.s. of each equation.

Equations (\ref{modellovecchio}) are obtained in a well known way,
by eliminating the field in the coupled equations for matter and field;
we however recall it  here  briefly  for the sake of completeness.
Working in the dipole aproximation means to linearize the system  with
respect to the displacements $q_j$ and to their time  derivatives.
 So the field ``created'' by particle $k$ is obtained from
the Maxwell equations by taking as sources the density  and the
current density given by 
\begin{equation}\label{corrente}
 \begin{array}{rcl} \rho_k(\vett x) &=& e \delta(\vett x -r_k \vett i)
    - e q_k\vett i \cdot \nabla \delta(\vett x -r_k \vett i) \\ \vett
    j_k(\vett x) &=& e \dot q_k\vett i \, \delta(\vett x - r_k \vett
    i) \ .  \end{array}
\end{equation}
respectively, $\delta$ being the usual ``delta function''. The first
source term  $e \delta(\vett x -r_k \vett i)$ in the density gives
rise to the static Coulomb field, which  was already taken into
consideration. One then solves 
the Maxwell equations with  the other terms in the
sources  (\ref{corrente}), and  the retarded fields thus   come out in
a natural way if  the solutions are evaluated,  as  usual, at a time
$t>t_0$ where $t_0$ is an ``initial time''; however, advanced fields 
too could have  been
used, by  matching the initial data through a suitable  free field.
The analytical computations leading to the retarded fields are classical,
and are easily reproduced. The results can however be  found
in reference \cite{becker} (see page 284). The magnetic field turns out to
be already of first order. So  the magnetic force is of second
order, and drops out, and  one remains with the electric force. The
electric  field naturally  appears as decomposed into the sum of several
terms, but due to the specific model considered here (displacements along a
line, on which the dipoles themselves are lying), some terms cancel and
other ones suitably add up, so that one remains with two terms only, which
are the ones appearing at the r.h.s. of equation (\ref{modellovecchio}). 
In conformity to the dipole approximation, they are evaluated  at the
equilibrium positions of the  particles.

We finally add a few words concerning the old problem of  the ``self--field'',
i.e.  the fact that the   electric field ``created'' by any
particle $j$ diverges at the position of that particle itself, so that
a suitable prescription is needed.  Here we follow the long tradition,
going back to Lorentz, Planck and Abraham, and definitely fixed by
Dirac (see also \cite{marino}),  
%%%%%%
according to which the force due the
self-field leads both to  mass renormalization (so that the
empirical mass $m$ is introduced as an external parameter) and to the
familiar radiation reaction force,  which yelds, in the dipole
approximation,  the term 
$-m\varepsilon\tdot q_j$ at the l.h.s. of equations (\ref{modellovecchio}).

\section{The Oseen identity  and the normal modes} 
So, our model is defined by the system of equations
(\ref{modellovecchio}). For an analytical discussion we have 
to  make  use of the 
 the relation $r_{jk}=|r_j-r_k|=|j-k|\, a$ holding in our specific 
one--dimensional model. With the spontaneous relabeling $k-j=n\in
\interi\setminus \{0\}$,
the final form of the equations to be studied here  is thus  (for $j\in\interi$)
\begin{equation}\label{modello}
 \ddot q_j + \omega_0^2 q_j -\vareps \tdot q_j =\fraz{ 2e^2}{ma^3}
        \sum_{n\ne 0}  \bigg[ \fraz {q_{j+n}(t-|n|a/c)}{|n|^3} + \fraz ac
        \fraz {\dot q_{j+n}(t-|n|a/c)}{|n|^2} \bigg] \ .
\end{equation}

The first point we make concerning system (\ref{modello}) is that it
cannot present damped normal modes, contrary to what might be expected
according to the generic presumption  that  charged particles,
when accelerated, should radiate energy away. In fact, if one looks
for normal modes, i.e.  introduces  the
\emph{ansatz} (the real part should actually be taken later)
\begin{equation}\label{ans1}
q_j(t)=u_j \exp(i\omega t) \ ,
\end{equation}
where the parameter $\omega$ is a priori a complex number, then 
(\ref{modello}) yelds the system of equations
\begin{equation}\label{modi}
 (-\omega^2 + \omega^2_0 + i\vareps \omega^3)\, u_j = \fraz
    {2e^2}{ma^3} \sum_{n\ne 0} u_{j+n} \exp
    \big[-i|n|a\omega/c\big]\left( \fraz 1{|n|^3} + i \fraz{a\omega}c
    \fraz 1{|n|^2} \right) \ .
\end{equation}
Now, if $\omega$ had a \emph{positive} imaginary part (which gives the
familiar damped solution, when dealing with one single dipole subjected
to an external forcing), then the terms of the series at the
r.h.s. would grow exponentially fast, and the series would diverge. If
instead $\omega$ had a \emph{negative} imaginary part, then one would
be in presence of a so called runaway solution, i.e.  a motion
$q_j=q_j(t)$ diverging for $t\to +\infty$. Following Dirac, Planck and
the other classical authors, runaway solutions are discarded on
physical grounds; and the same we also do. In more explicit terms, as
an essential part of the definition of our model  we restrict our
consideration to motions $q_j=q_j(t)$ that are solutions to  the
equations (\ref{modello}) and in addition satisfy the nonrunaway
conditions of being  bounded for all times.

So the problem of obtaining normal modes for our model is reduced to
finding solutions of the form (\ref{ans1}) to  system (\ref{modi}), with  $\omega$
\emph{real}.  To this end we introduce the  further usual \emph{ansatz}
\begin{equation}\label{ans2}
u_j=C\exp(i\kappa aj) \ ,
\end{equation}
with a given parameter (the ``wave number'') $\kappa\in [-\pi/a, \pi/a)$. 
This corresponds to considering a ``material wave'' with \emph{phase
velocity} $v=\omega/\kappa$.  So one is reduced to a single complex 
equation, namely
\begin{equation}\label{dispersione}
 -\omega^2 + \omega^2_0 + i\, \vareps \omega^3 = \fraz {2e^2}{ma^3}\,
        \bigg[ f(\kappa a, a\omega/c) + \,  i\,  g(\kappa
        a, a\omega/c) \bigg] \ ,
\end{equation}
where we have introduced the two functions 
\begin{eqnarray}\label{effe}
f(\alpha, \beta) &=& \sum_{n\ne 0} \bigg(\fraz{\cos(n\alpha-|n|\beta)}
             {|n|^3} -\beta \,
             \fraz{\sin(n\alpha-|n|\beta)}{|n|^2}\ \bigg)\\
\label{gi}
g(\alpha, \beta)&=& \sum_{n\ne 0} \bigg( \fraz{\sin(n\alpha-|n|\beta)}
                    {|n|^3} + \beta \, \fraz{\cos(n\alpha-|n|\beta)}
                    {|n|^2}\ \bigg)\ .
\end{eqnarray}
This corresponds to  two real equations, namely
\begin{equation}\label{dispersionereale}
-\omega^2+\omega_0^2=\fraz {2e^2}{ma^3}\,  f(\kappa a, a\omega/c)
\end{equation}
\begin{equation}\label{dispersioneim}
\varepsilon \omega^3=  \fraz {2e^2}{ma^3}\, g(\kappa a,a\omega/c) \ , 
\end{equation}
in the two
real variables $\omega$ and $\kappa$, with parameters $a$ and
$\omega_0$ (while $e$, $c$, $m$ and $\varepsilon$  are thought of as fixed).
So there should be no
possibility for the solutions to define implicitly a curve in  
the ($\kappa, \omega$) plane,
 typically a function $\omega=\omega(\kappa)$, as 
expected for a dispersion relation in an infinite lattice.
The situation turns out however to be quite fortunate,
because it can be established that the second equation
(\ref{dispersioneim})  actually is an
identity (which we like to call the Oseen identity), so that one
remains with only one equation, i.e. (\ref{dispersionereale}),  in two variables. 

This is established as follows.  While
the series $f$ entering the real part is not expressible in terms of
elementary functions, it occurs that the series $g$ entering the imaginary part can
be summed without pain. This amounts to establishing  the classical formulas
\begin{equation}\label{somma3}
\sum_{n=1}^{+\infty} \fraz {\sin (nx)}{n^3}=\fraz {x^3}{12}-\fraz{\pi
              x^2}4 +\fraz {\pi^2 x}6
\end{equation}
and
\begin{equation}\label{somma2}
\sum_{n=1}^{+\infty} \fraz {\cos (nx)}{n^2}=\fraz {x^2}{4}-\fraz{\pi
              x}2 +\fraz {\pi^2 }6 \ ,
\end{equation}
which are known to hold in the fundamental domain  $x\in [0,2\pi)$; see for example
 the handbook of Abramovitz and Stegun \cite{abramo}. 
In such a way one obtains 
\begin{equation}\label{cancellazione}
g(\alpha, \beta)= \left\{ \begin{array}{ll} \beta^3/3& \mbox{ if } |\beta/\alpha|<1
 \\ \beta^3/3 + \pi/2(\alpha^2 -\beta^2)   & \mbox{ if
 }|\beta/ \alpha| \ge 1 \ .\end{array} \right.
\end{equation}
The first of these  is proved  directly using  the formulas
 (\ref{somma3}), (\ref{somma2}),
 while the second one is established  by translating the
 variable $\alpha+\beta$ or the variable $\alpha-\beta$, when
 required, to  the fundamental domain $[0,2\pi)$.

Thus it turns out that, in the
domain of the ($\kappa$,$\omega$) plane where 
$|v|/c<1$  (which corresponds to $|\beta/ \alpha|< 1$),
 everything  combines in such a miracolous way that 
equation (\ref{dispersioneim}) rather turns out to be  an identity. In
such a domain, the dispersion relation  is then
defined implicitly by the real transcendental equation (\ref{dispersionereale}),
which is discussed in the next Section. It
is possible to check that in the complementary domain 
$|v|/c\ge 1$ there are  no further solutions to the
complex equation (\ref{dispersione}). The waves having the property $|v|/c<1$
are known in optics as evanescent waves.

\section{The dispersion relation}

We come now to a discussion of the dispersion relation, namely the
curve in the ($\kappa$,$\omega$) plane  implicitly defined by equation
(\ref{dispersionereale}), depending on the parameters
$a$ (the lattice step) and $\omega_0$ (the proper frequency of the
dipoles). The first thing to be established is whether there
are values of the parameters for which a curve in fact exists at all;
then one would like to determine some qualitative features, such as
for example whether the curve is the graph of a function
$\omega=\omega(\kappa)$, and how it differs from the constant function
$\omega(\kappa)=\omega_0$.

An analytical study is actually nonexpedient, because the function $f$
entering equation (\ref{dispersionereale}) turns out to be, at variance with $g$,
non expressible in terms of elementary functions. So we turn to a
numerical study, taking a pragmatic attitude. We have to look for
possible intersections of the parabolid $z=-\omega^2+\omega_0^2$ with
the surface $z=({2e^2}/{ma^3})\,\,  f(\kappa a,\omega a/c)$, by  approximating
the function $f$ through  a suitable truncation of the series (\ref{effe}).
 
The parameters $a$ and $\omega_0$ could a priori be taken each  in the whole
positive axis, but here we limit ourserlves to the consideration of
some  values having an order
of magnitude of interest for atomic physics. Actually, for $\omega_0$
we just  consider one value, i.e.  the rotational frequency of the electron
in the hydrogen atom in circular motion at the Bohr radius $R_B$;
for $a$ we take several values ranging from $0.1\, R_B$ up to $5\, R_B$.
The results are illustrated in Fig. 1,  where the dispersion curve is 
reported for several values of $a$ (indicated in the Figure in units of $R_B$).

The most significant result seems to be that
 the dispersion curve indeed exists; moreover its topology depends on
 the value of the
 lattice step $a$. All curves have a common behavior at the right extreme of
 the Figure, because  they all
 intercept the vertical line $\kappa a=\pi$
 with a horizontal tangent. The situation is instead different in the
 region of  small $\kappa$. Indeed,  there exists a critical value
 $a^*\simeq 1.7\,  R_B$ of $a$.
For $a>a^*$ the curve is the graph of a function $\omega=\omega(\kappa)$ 
(in the whole admissible domain of $\kappa$) which, for 
 increasing  $a$,  tends to the  horizontal curve
 $\omega=\omega_0$; actually, the function essentially coincides with the 
constant function $\omega(\kappa)=\omega_0$
already for $a\simeq 5\,  R_B$. Instead, for $a<a^*$ the curve is the
 graph of a function $\kappa=\kappa (\omega)$,
 which has a central part tending to  to the vertical curve
 $\kappa a=\pi/2$  as $a$ decreases. 
Notice that in the ($\kappa$, $\omega$) plane the curves can exist only
below the line $\omega/\kappa=c$. Such a line is reported in the
 Figure for the case $a=5\, R_B$. Notice that the slope increases as
 $a$ diminishes, so that eventually the line becomes indistinguishable
 from the axis of the ordinates; in the Figure,
 this  would already  occur fo $a=R_B$.

\begin{figure*}
\includegraphics[width=11cm]{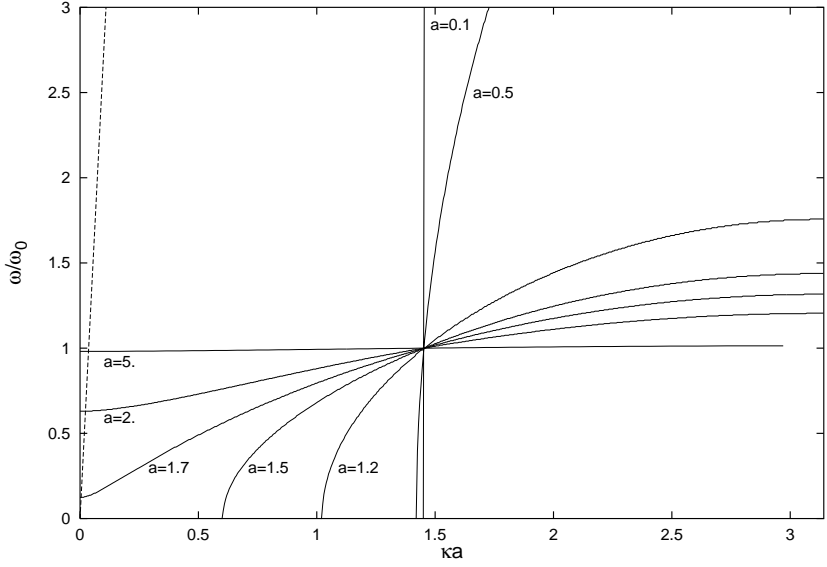}% Here is how to import EPS art
\caption{\label{fig:1} The dispersion curves in the plane 
$(\kappa a,\omega/\omega_0)$, for some values of $a$.}
\end{figure*}

\section{The Oseen identity in terms of the forces}

It has been shown in Section 3 how the Oseen identity  (namely the
identically vanishing of the imaginary part of the complex equation 
(\ref{dispersione}) for the normal modes)
allows for the
existence of a dispersion relation.  We now investigate how the identity reads
 in terms of the forces acting on each dipole. We  show the quite
significant result that such an identity provides
a cancellation of the radiation reaction term $-m\varepsilon \tdot
q_j$ pertaining to any dipole  $j$ by a resummation of a part of the retarded
forces due to all the other
dipoles $k\ne j$. In the next Section  we will  show that the
identity can 
be expressed in another very enlighting way, which will allow us to make a strong
connection with the work of Wheeler and Feynman.

Let us 
rewrite the equations of motion of our model in a perhaps more
transparent way as follows:
\begin{equation}\label{dirac}
m (\ddot q_j+\omega_0^2 q_j-\varepsilon \tdot q_j)= e \sum_{k\ne j} 
E_{jk}^{\mathrm{ret}}\ .
\end{equation}
Here $E_{jk}^{\mathrm{ret}}$ is the (component along the $\vett i$
vector of the) retarded electric field ``created'' by particle $k$
and evaluated  at the equilibrium position of particle $j$, in the
dipole approximation, namely:
\begin{equation}\label{ret} 
E_{jk}^{\mathrm{ret}}= 2e\, \bigg[ \fraz {q_k(t-r_{jk}/c)}{r_{jk}^3} 
          + \fraz 1c\,  \fraz {\dot q_k(t-r_{jk}/c)}{r_{jk}^2}\bigg]  \ .
\end{equation}
Now, looking back at the way in which the existence of normal modes
was proved, it is obvious that the result found in Section 3 can 
equivalently be expressed in the following way: There exist
 normal--mode solutions  $q_j(t)= A \cos (\kappa a
j-\omega t)$ of system (\ref{dirac}) (\ref{ret})  such that the sum of the 
retarded forces acting on  any dipole $j$ due to all other dipoles
$k\ne j$ decomposes as
\begin{equation}\label{forzaj}
e \sum_{k\ne j} E_{jk}^{\mathrm{ret}}= \fraz {2e^2}{a^3} f(\kappa a,
a\omega/c) \, q_j(t)\, -m\, \varepsilon\,  \tdot q_j (t)   \ ,
\end{equation}
i.e. into a term that exactly compensates  the ``radiative term'' at the l.h.s. 
of (\ref{dirac}), and into another one that  corrects the mechanical
frequency $\omega_0$ as to have $\ddot q_j+\omega^2 q_j=0$ (as
obviously should be, by the definition itself of a normal--mode solution).

One could now ask whether it is possbile
to describe in some more
trasparent way such a  splitting of the sum of the retarded forces acting on
dipole  $j$  into a  part compensating the radiative term, and
another one  correcting the mechanical frequency $\omega_0$. This
is actually the  door through  which  the advanced forces naturally enter the
arena, and one is somehow compelled to take them into consideration,
notwithstanding the fact that, following the traditional approach, only
 retarded forces had  originally been introduced in the model.

Indeed, the above decomposition of the total retarded force acting on
dipole $j$ into a term proportional to $q_j$ and another one
proportional to $\tdot q_j$ 
turns out to actually constitute a decomposition into a 
symmetrical part and  an 
antisymetrical one  with respect to time reversal. On the other hand the most 
natural decomposition of such a type for the single  retarded forces
themselves
 is  nothing but 
\begin{equation}\label{somma}
E_{jk}^{\mathrm{ret}}=
\fraz{E_{jk}^{\mathrm{ret}}+E_{jk}^{\mathrm{adv}}}2 \, +
\fraz{E_{jk}^{\mathrm{ret}}-E_{jk}^{\mathrm{adv}}}2 \ ,
\end{equation}
where $E_{jk}^{\mathrm{adv}}$  is the  advanced field,
which, in our case,  in the  dipole approximation reads
\begin{equation}
E_{jk}^{\mathrm{adv}}=2e\, \bigg[ \fraz {q_k(t+r_{jk}/c)}{r_{jk}^3} -  
                  \fraz 1c \fraz {\dot  q_k(t+r_{jk}/c)}{r_{jk}^2} \bigg]\ .  
\end{equation}
So the semidifference of the retarded and the advanced forces has the expression
\begin{eqnarray}
e\, \sum_{k\ne j} \fraz {E_{jk}^{\mathrm{ret}} - E_{jk}^{\mathrm{adv}}}2 
   &=&   2 e^2 \sum_{k\ne j} \bigg[ \fraz
   {q_k(t-r_{jk}/c)-q_k(t+r_{jk}/c)}{2 r_{jk}^3}  +
        \nonumber \\
   &+& \fraz 1c \fraz {\dot q_k(t-r_{jk}/c) + \dot q_k(t+r_{jk}/c)}{2
   r_{jk}^2} \bigg] \ ,
\end{eqnarray}
and it is immediately checked,  using the Oseen identity,  that along
any  normal--mode solution one has
 \begin{equation}\label{oseen2}
e\sum_{k\ne j} \fraz {E_{jk}^{\mathrm{ret}} - E_{jk}^{\mathrm{adv}}}2 
=-\, m\varepsilon\, \tdot q_j \ . 
\end{equation}

Due to the linearity of the equations of motion, this result can be
extended to any combination of linear modes, and so one is lead
to  the main result of the present Section, namely:  In
a generic solution of  system (\ref{dirac}) (\ref{ret}),
the ``reaction radiation force'' acting on each dipole is exactly
 compensated by the sum of the  semidifferences
of the retarded and the advanced forces due to all the other dipoles, 
i.e. the relation (\ref{oseen2}) holds.

So we have the following situation. In the original definition of our
model, the force acting on
particle $j$ had been  defined, in the familiar way,  as the
Lorentz force (in the dipole approximation) due to the electromagnetic
field. By the standard procedure of  eliminating the field in the
coupled  equations of matter and field, such a force was then 
represented as  the sum of the retarded forces
 due to all other particles $k\ne j$. On the other hand, such a
resultant  retarded force  can be looked upon as being 
split into the combination (\ref{somma})  of the semisum and
the semidifference of the retarded and the advanced forces due to all
the other particles. But  the
Oseen identity in the form (\ref{oseen2}) then shows that the  
semidifferences just add up  in such a way as to exactly cancel the
reaction radiation term pertaining to particle $j$. 

This has the important consequence that in the original system
defining the model the radiation reaction term appearing in each equation can be
dropped, provided that  the r.h.s.  be changed in a corresponding way, 
namely  with  each retarded force  replaced by the corresponding
semisum of  retarded and advanced forces. So the original system of
equation  (\ref{dirac}) can equivalently be rewritten in the form
\begin{equation}\label{sf} 
m\, (\ddot q_j+\omega_0^2 q_j) 
= e\,  \sum_{k\ne j}\,  
\fraz {E_{jk}^{\mathrm{ret}}+ E_{jk}^{\mathrm{adv}}}2 \ .
\end{equation}

\section{The Oseen identity as a weak form of the Wheeler--Feynman identity}

We now rewrite the Oseen identity (\ref{oseen2}) in a more perspicuous
form. To this end we have  to introduce the quantity
$\big[ {E_{jj}^{\mathrm{ret}}-E_{jj}^{\mathrm{adv}}}\big]/2$.
Apparently, this  is not defined, inasmuch  as it involves two diverging 
terms. However, one immediately sees that such singularities are removable,
that the quantity is correctly defined, and in fact one has
\begin{equation}\label{Ejj}
e\, \fraz {E_{jj}^{\mathrm{ret}}-E_{jj}^{\mathrm{adv}}}2 
=m \varepsilon \tdot q_j\ .
\end{equation}
Indeed the actual original quantities of interest are the fields
$E_{(k)}^{\mathrm{ret}}(\vett x)$ and $E_{(k)}^{\mathrm{adv}}(\vett
x)$ ``created''  by particle $k$ and evaluated at the current point
$\vett x$, because  the quantities entering the model are nothing but 
such fields evaluated at the equilibrium position $\vett r_j$ of
particle $j$, i.e.  $E_{jk}^{\mathrm{ret}}= 
E_{(k)}^{\mathrm{ret}}(\vett r_j)$, and the 
corresponding advanced quantity.  Now,   evidently 
$E_{(j)}^{\mathrm{ret}}(\vett x)$ diverges as $\vett x\to
\vett r_j$, but from the explicit expression one immediately checks
that the limit exists  for the semidifference, and that its value is
given according to (\ref{Ejj}). This in fact just is  a
particular case of a general result found  by Dirac.

The conclusion is that, in virtue of (\ref{Ejj}), the Oseen
identity (\ref{oseen2}) now reads 
\begin{equation}\label{oseen3}
\sum_{k\in\interi}\fraz {E_{jk}^{\mathrm{ret}}-
E_{jk}^{\mathrm{adv}}}2\,  =0\ ,  \qquad j\in\interi \ . 
\end{equation}

We now come to the connection with the work of  Wheeler and
Feynman \cite{wf}. The authors  point out
that there exist two a priori different formulations of
electrodynamics of point particles, namely what they call ``\emph{the
theory of Schwarzschild and Fokker}'' on the one hand, and the `` \emph{theory of
Dirac}'' on the other. The latter, which is the traditional one, includes the radiation reaction
term $-m\varepsilon \tdot q_j$ and introduces   retarded forces; 
the first one drops the radiation reaction term
and introduces  the semisum of the retarded and the  advanced forces.
  In our model, such theories  amount to nothing
but   equations (\ref{sf}) and (\ref{dirac}) respectively.  The
declared  aim of Wheeler and Feynman (see page 170 of their paper)
was to prove ``\emph{a complete equivalence between the theory of
Schwarzschild and Fokker on the one hand and the usual formalism of 
electrodynamics} (i.e. that of Dirac) \emph{ on the other}''. 

They are able to prove the equivalence by  making use of an hypothesis, which
they describe in physical terms as corresponding to the existence of
an ``\emph{absorbing universe}''. In mathematical terms such an hypothesis
is formulated as requiring the identically vanishing of the
semidifference of the fields created by all the particles, i.e. the identity 
\begin{equation}\label{wf}
 \sum_{k\in\interi}\fraz {E_{(k)}^{\mathrm{ret}}(\vett x)-
E_{(k)}^{\mathrm{adv}}(\vett x)}2\, =0\ , \qquad \vett x\in\reali^3 \ . 
\end{equation}
In fact, for the equivalence it is sufficient to assume that the above relation
holds just  at the  positions of all the particles and not in the full space
$\reali^3$. Now, in our model we have shown that the identity
in the latter weaker form is not an additional hypothesis, but rather
a theorem. Thus the equivalence of the two  formulations of electrodynamics
of point particles according to Schwarzschild--Fokker  
and to  Dirac is proven in our model. 

We finally add  a comment, concerning the way in which Wheeler and Feynman
discuss the equivalence of the two formulations. They give four
arguments, with headings ``\emph{The radiative reaction: derivation
I,II,III,IV}''. The fourth ``derivation''  is essentially the one given here  in 
Section 5 (apart from the fact that they take as  a  postulate the
identity which we prove).  
On the other hand, in the previous ``derivations''  they
attempt essentially at proving (instead of  postulating) what we called
the Oseen identity in its first form (\ref{oseen2}). So,  it will not
appear strange that we happened to  understand the whole paper of
Wheeler and Feynamn, and in particular their ``derivations'',  only
after we proved ourselves the Oseen identity in our model.
The conclusion is thus  that we prove in a special model what 
they argument on general grounds. 
Conversely, this seems to be a strong indication that the
Oseen identity might be proved, as a real theorem, for a much larger
class of models.

\section{Conclusions}

So we have proven, at least for our particular linearized  model of 
many--body matter
radiation interaction, that  there exist nonradiating normal modes,
i.e. solutions to the equations of motion of the complete system
particles plus field  in
which the mechanical energy of the particles remains
constant, notwithstanding the fact that all particles perform
accelerated motions. A preliminary analytical investigation shows that the
same phenomenon occurs in  a
different  model,  in which, at variance with the present one, 
 also the far fields decaying as $1/r$ play a role.
A natural guess seems to be  that the same should  occur with a three
dimensional crystal. What should occur  for a disordered system
or a gas, is instead, apparently,  completely open. 
Another open question concerns the Wheeler--Feynman identity
(\ref{wf}). Indeed    the Oseen
identity  was shown here to be equivalent 
to a weak form of it, and  thus naturally the question arises
  whether the Wheeler--Feynman identity itself,
in its general form (\ref{wf}), holds.

We add now a further comment,  concerning 
the connection of the present work with   the
problem of a microscopic foundation of optics, especially  
for the  theories of dispersion and of extinction of light.
At first sight one might be tempted to believe  that the 
handbook of Born and Wolf did
already say the last word, at least for what
concerns the general aspects of the problem. But an accurate analysis
shows that actually  this is not the case. Indeed they do not deal with a
clearly defined mathematical model,  and somehow oscillate  between a
continuum phenomenological description of matter on the one hand, and
the consideration of single dipoles on the other; moreover, they do
not even introduce  an actual  dynamical equation for the dipoles.
According to Born and Wolf, the dynamical
foundation was given by Ewald \cite{ewald} and by  Oseen 
(in his first paper \cite{oseen1}).
Now, in our opinion no one of these  two works is
consistent. Indeed they both neglect the radiation reaction term, and
nevertheless  pretend that  normal modes do exist. But we have shown
that, at least for
a one--dimensional crystal, a part of the retarded forces acting on any given
dipole due to all the other ones, just add up in such a way as to 
 produce the  ``dissipative'' term $-m\varepsilon \tdot q_j$, which would
exactly compensate the reaction
radiation term, if this had originally been included  in the model. This  
is indeed the
reason for the very existence of normal modes in our model. But
conversely,  just for the same reason,  
normal modes cannot exist if the reaction radiation term had not
been included in the model.  In our opinion, there should be some mistake
hidden in the two quoted
works. It seems to us that the new relevant step after such works is just the one
performed  by Oseen in his subsequent paper \cite{oseen2}, where the idea of the
cancellation was introduced. Now, it happened that this second work of
Oseen was finally discarded as  wrong, and his proposal forgotten.
On the other hand,  the cancellation is    
proven    here as  a theorem in a
concrete model. In our opinion, the status of the  microscopic
foundation of optics, which should lead to an explanation of the
dispersion and the extinction of
light in molecular terms, should perhaps  be reconsidered.

\end{document}